\documentclass[a4paper,preprint,12pt]{revtex4}

\draft
\begin{document}


\title{Phase diagram of Pb(Zr,Ti)O$_3$ solid solutions from first principles}
 
\author{Igor A. Kornev$^{1}$, L. Bellaiche$^{1}$, P.-E. Janolin$^{2}$, B. Dkhil$^{2}$, and E. Suard$^{3}$}

\address{$^{1}$Physics Department,
                University of Arkansas, Fayetteville, Arkansas 72701, USA\\
                $^{2}$ Laboratoire Structures, Propri\'et\'es et Mod\'elisation des Solides, Ecole Centrale Paris, CNRS-UMR8580, Grande Voie des Vignes, 92295 Ch\^atenay-Malabry Cedex, France\\
                $^{3}$ Institut Laue-Langevin, 6 rue Jules Horowitz, BP 156 38042, Grenoble Cedex, France}

\date{\today}

\begin{abstract}
A first-principles-derived scheme, that incorporates ferroelectric and antiferrodistortive degrees of freedom, 
is developed to study finite-temperature properties of Pb(Zr$_{1-x}$Ti$_{x}$)O$_{3}$
solid solutions near its morphotropic phase boundary. The use of this numerical technique (i) resolves controversies about 
the monoclinic ground-state for some Ti compositions, (ii) leads to the discovery of an overlooked phase, and (iii) yields three multiphase points, that are each associated with four phases. Additional neutron diffraction measurements  strongly support some of these predictions.

\end{abstract}


\maketitle


\narrowtext

 The ferroelectric Pb(Zr$_{1-x}$Ti$_{x}$)O$_{3}$ (PZT) system is an example of perovskite solid solutions that are 
 of high technological relevance because of their widespread use in piezoelectric transducers and actuators \cite{Uchino}. 
 Its phase diagram in a narrow compositional region centered around $x$=0.50, and denoted as the morphotropic phase 
 boundary (MPB) \cite{Lines}, also makes PZT of large fundamental importance. For instance, recent measurements  have discovered an unexpected ferroelectric (FE) monoclinic phase  \cite{Noheda1} - that leads 
 to high electromechanical coefficients \cite{Beatriz,PRLPZT} and acts as a
structural bridge between the well-known FE tetragonal and rhombohedral phases  \cite{Noheda1,PRLPZT}.  Moreoever,  another 
overlooked monoclinic phase has been recently observed  \cite{Noheda2002,Hatch,Woodwoad,Ranjan}, in which  the 
usually-competing \cite{DaviZhong1998} ferroelectric distortions  and antiferrodistortive (AFD) motions (that are associated with 
the oxygen octahedra rotation) {\it coexist}. However,  many controversies still surround this latter monoclinic phase, such as 
its exact space group, the axis about which the oxygen octahedra rotate and even if it is a minority phase
rather than the long-range ground-state  \cite{Noheda2002,Hatch,Woodwoad,Ranjan}. 

In view of these findings, one may also wonder if  {\it other} surprises -- such as other  overlooked phase(s) or multiphase points 
(that are unusual thermodynamic features \cite{Multiphase}) --
can still be  in store for PZT. Some features could indeed have been missed because of (i)  the  difficulties 
in growing and characterizing at various temperatures many samples with tiny compositional differences; and (ii) the current lack of 
theoretical tools that are able to accurately compute finite-temperature properties of perovskite alloys. 
For instance, the two sole first-principles-based finite-temperature schemes \cite{PRLPZT,Grinberg} that yield a monoclinic phase 
in the phase diagram of disordered PZT have some shortcomings: the alloy effective Hamiltonian model of Ref.  \cite{PRLPZT} gives Curie 
temperatures that are $\simeq$ 60\% larger than measurements and does not include AFD motions, while we are not aware of any study reporting the 
accuracy of  the bond-valence model of Ref. \cite{Grinberg} for transition temperatures in PZT.

The purpose of this Letter are three-folds: (1) to demonstrate that it is possible to develop a numerical scheme able to accurately 
compute finite-temperature properties of perovskite solid solutions and that incorporates both FE and AFD motions; (2) to apply such scheme to resolve the controversies 
discussed above, as well as, to reveal some striking overlooked effects in the MPB of PZT; and (3) to experimentally confirm some of these discoveries, via neutron diffractions.

Our numerical scheme is based on the generalization of the first-principles-derived alloy effective Hamiltonian of Ref.  \cite{PRLPZT} to include AFD effects, in addition to FE degrees of freedom. The total energy is thus 
written as a {\it sum} of two main terms, $ E_{\rm FE}  (\{ { \bf u_{\it i}} \}, \{ \eta_{\it H} \}, \{  \eta_{\it I} \},  \{ \sigma_{\it j} \})$ and $E_{\rm AFD-C} (\{ { \bf u_{\it i}} \}, \{ \eta_{\it H} \}, \{  \eta_{\it I} \},  \{ \sigma_{\it j} \}, \{ { \bf \omega_{\it i}} \})$, 
where $E_{\rm FE}$ is the energy provided in Ref. \cite{PRLPZT}, while 
$E_{\rm AFD-C}$ is an additional term gathering AFD motions and their couplings in solid solutions.
${\bf u_{\it i}}$ is the local soft mode (which is directly proportional to the dipole) in unit cell $i$; $ \{ \eta_{\it H} \}$ 
and $\{  \eta_{\it I} \}$ are the homogeneous and inhomogeneous strain tensors \cite{ZhongDavid}, respectively;  $\sigma_{\it j}$=+1 or $-1$ corresponds to the presence of a Zr or Ti atom, 
respectively, at the B-lattice site $j$ of the Pb(Zr$_{1-x}$Ti$_{x}$)O$_{3}$ system.  Finally,  $\{ { \bf \omega_{\it i}} \}$ is 
a (B-centered) vector characterizing the direction and magnitude of the AFD motions in unit cell $i$.  For instance, 
${ \bf \omega_{\it i}} =0.1 {\bf z}$ corresponds to a rotation of the oxygen octahedra by 0.1 radians about the z-axis.  
For  $E_{\rm AFD-C}$, we propose the following expression that contains five major terms:
\begin{eqnarray}
  E_{\rm AFD-C} (\{ { \bf u_{\it i}} \}, \{ \eta_{\it H} \}, \{  \eta_{\it I} \},  \{ \sigma_{\it j} \}, \{ { \bf \omega_{\it i}} \}) = 
   \sum_i  [ \kappa_{A} \omega_{\it i}^2 + \alpha_{A} \omega_{\it i}^4 + \gamma_{A} (\omega_{\it i,x}^2 \omega_{\it i,y}^2 
   + \omega_{\it i,y}^2 \omega_{\it i,z}^2 + \omega_{\it i,x}^2 \omega_{\it i,z}^2) ]  \nonumber \\
    + \sum_{i,j} \sum_{\alpha,\beta} K_{i j,\alpha \beta} ~\omega_{\it i,\alpha} \omega_{\it j, \beta}     +   
    \sum_i \sum_{\alpha,\beta} C_{l ,\alpha \beta}~ \eta_{l}(i) ~\omega_{i, \alpha}  \omega_{i, \beta}      \nonumber\\ 
     +   \sum_i \sum_{\alpha,\beta,\gamma,\delta} D_{\alpha \beta \gamma \delta} ~\omega_{i, \alpha} \omega_{i, \beta}  
     u_{i, \gamma}  u_{i, \delta} + \sum_i  \omega_{\it i} [ A_0 \sigma_{\it i} +\sum_j A_1  \sigma_{\it j}] \nonumber \\
     \;\;
\end{eqnarray}

where the sums over $i$ run over all B-sites while the sums over $j$ run over the six B nearest neighbors of the B-site $i$. 
$\eta_{l}(i)$ is the $l^{th}$ component, in Voigt notation, of the total strain (i.e., homogeneous and inhomogeneous parts) 
at the site $i$. $\alpha$, $\beta$, $\gamma$ and $\delta$  denote Cartesian components, with the x-, y- and z-axes being along the 
pseudo-cubic [100], [010] and [001] directions, respectively.
The first four terms of Eq.~ (1) characterize  energertics of  
the hypothetical {\it simple} Pb$<B>$O$_{3}$ system resulting from the use of the virtual crystal 
approximation (VCA) \cite{LaurentDavid3}  (via averaging of the Zr and Ti potentials) to mimic 
Pb(Zr$_{0.5}$Ti$_{0.5}$)O$_{3}$, while the last term represents how the actual atomic distribution  affects the 
AFD motions. The first four terms are the self-energy associated with AFD motions, the short-range interaction energy between AFD degrees 
of freedom, the interaction energy between strain and ${ \bf \omega_{\it i}}$, and the interaction energy between local-modes and AFD motions, 
respectively. The $\kappa_A$, $\alpha_{A}$ and $\gamma_{A}$ parameters, as well as the $K_{i j,\alpha \beta}$, $C_{l ,\alpha \beta}$ and
 $D_{\alpha \beta \gamma \delta}$ matrices (that greatly simplify due to symmetry), are determined by performing local-density-approximation 
 (LDA) calculations \cite{LaurentDavid3,LDA} on 10 atom-supercells of Pb$<B>$O$_{3}$. On the other hand, the $A_0$ and $A_1$ parameters are 
 evaluated from LDA calculations on supercells containing {\it real} Zr and Ti atoms in addition to VCA atoms. 
Note that the proposed  expression for $E_{\rm AFD-C}$ differs from the one of Ref.~\cite{DaviZhong1998} for 
simple perovskites by its analytical expressions and the incorporation of alloying effects. 

The total energy of our effective Hamiltonian is  
used in Monte-Carlo (MC) simulations to compute finite-temperature properties of PZT systems.  
We use $12\times 12\times 12$ supercells (8640 atoms) and up to 4 million MC sweeps, as well as  decrease the 
temperature in small steps, to get well-converged results. 
The $\{ \sigma_{\it j} \}$ variables are chosen randomly to mimic disordered alloys. 
Outputs of the MC procedure  identify the space group of the predicted phases, and 
are: the homogeneous strain tensor $\{ \eta_{\it H} \}$, the  $< {\bf u}>$ supercell
average of the  local mode vectors $\{ { \bf u_{\it i}} \}$, and the  $< {\bf \omega}>_{R}$ vector defined as 
$< {\bf \omega}>_{R} = \frac{1}{N} \sum_{i} { \bf \omega_{\it i}} (-1)^{n_x(i)+n_y(i)+n_z(i)}$, where the sum runs over  
the N sites $i$ and where $n_x(i)$, $n_y(i)$ and $n_z(i)$ are integers locating the cell $i$  (i.e.,  the B-site $i$ is centered at $[n_x(i) {\bf x}+n_y(i) {\bf y} +n_z(i) {\bf z}] a$, where $a$ is the lattice constant of
PZT and where ${\bf x}$, ${\bf y}$ and ${\bf z}$ are unit vectors along the Cartesian axes). A non-vanishing $< {\bf u}>$ indicates ferroelectricity while a non-zero $< {\bf \omega}>_{R}$ characterizes AFD motions associated with the R-point of the cubic first Brillouin zone. We also use the correlation functions of Refs~\cite{Alberto2}  to precisely determine  transition temperatures. 

Figure 1(a) shows the x-, y- and z-Cartesian
coordinates ($<u_{1}>$, $<u_{2}>$ and $<u_{3}>$) of  $< {\bf u}>$ in disordered Pb(Zr$_{0.52}$Ti$_{0.48}$)O$_3$ as a function 
of temperature, as predicted by the alloy effective Hamiltonian approach of Ref.  \cite{PRLPZT} {\it  that does not include AFD effects}. 
Figures 1(b) and 1(c) display $< {\bf u}>$ and $< {\bf \omega}>_{R}$, respectively, in the same material, but when using the present method -- with $<\omega_{1}>_R$, $<\omega_{2}>_R$ and $<\omega_{3}>_R$ denoting the x-, y- and z-Cartesian coordinates of $< {\bf \omega}>_{R}$.
Figure 1(a) indicates that neglecting AFD motions and its couplings with ferroelectricity and strain leads to (1) a transition from a cubic 
paraelectric ${\it Pm}\bar{3}{\it m}$ phase (in which  $< {\bf u}>$ is close to zero) to a tetragonal {\it P}4{\it mm} ferroelectric phase (in which $<u_{3}>$ 
drastically  increases while $<u_{1}>$ and $<u_{2}>$ remain nearly null) around 825K $\pm$ 5K -- which is higher than the experimental Curie 
temperature of $\simeq$ 663 $\pm$ 5 \,K \cite{Noheda2000} ; and (2) a tetragonal-to-monoclinic {\it Cm}  transition, for which $<u_{1}>$ and $<u_{2}>$ 
now increase and become equal to each other while still being smaller in magnitude than  $<u_{3}>$ , at 250K $\pm$ 13K. 
Comparing  Figs 1(b-c)   with Fig~1(a) reveals that turning {\it on} AFD motions has dramatic effects on the phase transition sequence -- as correctly guessed by Ref. \cite{Fornari}. As a matter of fact, two new phases -- with both FE and AFD components -- are found at the lowest temperatures, namely monoclinic {\it Cc} between $\simeq$ 163 \,K 
and $\simeq$ 138 \,K and rhombohedral {\it R}3{\it c} below 138\,K $\pm$ 6\,K.
Furthermore, the theoretical Curie temperature is decreased down to 413 \,K $\pm$ 13 K, which now makes it smaller than the measurement \cite{Noheda2000}. This discrepancy is due to the fact that the effective Hamiltonian parameters 
were determined at the (underestimating) theoretical LDA lattice constant. 
As a matter of fact, we numerically found that the pressure leading to our theoretical volume matching the experimental one \cite{Noheda2002} at low temperature is $-$4.68 GPa, and that the use of this
negative pressure within the present numerical scheme results in  the $< {\bf u}>$ and $< {\bf \omega}>_{R}$ displayed in Figs~1(d) 
and (e), respectively. The resulting theoretical Curie temperature is now $\simeq$ 650 $\pm$ 13\,K, which is in excellent agreement with the 
measured value of $\simeq$ 663 $\pm$ 5 \,K. Moreover,  the symmetry of the purely FE phases coincides with those measured at this 
specific composition \cite{Noheda2002}, that is  tetragonal  {\it P}4{\it mm} and monoclinic {\it Cm} when decreasing the temperature. Our simulations also predict 
the existence of a monoclinic state below 138\,K $\pm$ 13\,K in which ferroelectricity cohabits with AFD motions, as consistent with the  
electron-diffraction reflections observed in Ref. \cite{Noheda2002}  below 150\,K. Furthermore, we confirm the findings of Refs 
\cite{Noheda2002,Hatch,Woodwoad} that the correct space group for the low-temperature phase of Pb(Zr$_{0.52}$Ti$_{0.48}$)O$_3$ is {\it Cc} rather 
than {\it Pc}, as initially proposed in Ref. \cite{Ranjan}. Moreover, this {\it Cc} phase is indeed the (long-range-ordered) ground-state of 
Pb(Zr$_{0.52}$Ti$_{0.48}$)O$_3$ rather than a minority phase coexisting with the purely FE {\it Cm} phase for the lowest temperatures, as 
suggested in Ref. \cite{Noheda2002}.
Finally, the oxygen octahedra in this {\it Cc} phase do {\it neither} rotate about the [001] direction (as thought in Ref. \cite{Ranjan}) nor 
about the [111] direction (as proposed in Ref. \cite{Noheda2002}) but rather about an axis that is in-between the [001] and [111] directions
 (which is in agreement with Ref. \cite{Woodwoad}), since Fig~1(e) shows that, at the smallest temperatures,  $<\omega_{3}>_R$  is 
larger in magnitude than $<\omega_{1}>_R$ and $<\omega_{1}>_R$= $<\omega_{2}>_R$. Interestingly, this axis of rotation does {\it not} precisely coincide 
with the direction of the polarization - since the ``FE'' ratio $\frac{<u_{3}>}{<u_{2}>}$ differs from the ``AFD'' ratio $\frac{<\omega_{3}>_R}{<\omega_{2}>_R}$ -- 
which contrasts with the assumption of Ref. \cite{Woodwoad}.

We now turn our attention to Fig.~2 that displays the phase diagram of Pb(Zr$_{1-x}$Ti$_{x}$)O$_3$  near the MPB, as predicted by the present scheme with a $-$4.68 GPa pressure. Seven phases exist within a narrow compositional range. 
Five of such phases are  well-known, namely paraelectric ${\it Pm}\bar{3}{\it m}$, FE rhombohedral {\it R}3{\it m},  FE tetragonal  {\it P}4{\it mm}, FE monoclinic {\it Cm}, and the rhombohedral {\it R}3{\it c} state in which ferroelectricity coexists
 with AFD motions. The sixth phase is the previously ``controversial'' {\it Cc} phase 
 discussed above, and that is the ground-state at  intermediate Ti compositions. (Note that the monoclinic phases are predicted to exist for the 
 narrow Ti compositional range of 46.3 \%--51.5\%, which agrees remarkably well with the experimental range of  $\simeq$ 46\%-52\%  \cite{Noheda2000,Lima}.)
The last phase occurs at small temperature and for the largest displayed Ti compositions, and  
 has never been 
suggested to be a possible ground-state of PZT (to the best of our knowledge). This phase has a  {\it I}4{\it cm} space group, 
also involves a coexistence of ferroelectricity and 
rotation of oxygen octahedra, but is associated with the {\it tetragonal} symmetry (unlike {\it R}3{\it c} and {\it Cc}). As a result, an 
increase in Ti concentration from  $\simeq$ 47\% to 52\% results not only in a spontaneous polarization continuously rotating from [111] to [001] 
but also to the oxygen octahedra varying their rotation axis from [111] to [001] -- via the [uuv] directions in the bridging monoclinic 
structures {\it Cm} and {\it Cc} phases. [We also checked that {\it no} AFD state associated with an off-center k-point {\it different} from R exists in the MPB of PZT.]
 Other features of Fig.~2 going against common beliefs \cite{Noheda2002,Woodwoad} are the {\it non-verticality}  of the 
temperature-composition boundary lines separating rhombohedral from monoclinic structures. As a result, {\it five} different phases 
emerge for some narrow compositions. For instance, 
for x $\simeq$ 47.2\%, our scheme predicts the ${\it Pm}\bar{3}{\it m}$,  {\it P}4{\it mm}, {\it Cm}, {\it R}3{\it m} and {\it R}3{\it c} phases when decreasing the temperature.  Similarly, a Ti 
concentration around 47.5\% yields 
a ${\it Pm}\bar{3}{\it m}$ --  {\it P}4{\it mm} -- {\it Cm} -- {\it Cc} -- {\it R}3{\it c} sequence \cite{footnoteaccuracy}. 
Finally, Fig.~2 further reveals the existence of 
 three temperature-compositional points, for which {\it four} phases are 
very close to each other! (Multiphase points are unusual thermodynamic features 
\cite{Multiphase}.)
 The first multiphase
point gathers the paraelectric ${\it Pm}\bar{3}{\it m}$ state and the FE {\it R}3{\it m}, {\it Cm} and  {\it P}4{\it mm} phases, and occurs for $T$ $\simeq$ 610\,K and $x$ $\simeq$ 46.3 \% \cite{problemTc}. (Note that a quadruple point in which the paraelectric state coexists with three FE phases was also predicted to exist in the phase
diagrams of epitaxial PZT thin films \cite{Pertsev}.)
Two purely FE phases meet with  two phases exhibiting a coexistence of  FE and  AFD motions for the other two multiphase points. More precisely, 
{\it R}3{\it m}, {\it Cm}, {\it R}3{\it c} and {\it Cc} all meet at $T$ $\simeq$ 150 \,K and $x$ $\simeq$ 47.4\%, while 
{\it Cm},  {\it P}4{\it mm}, {\it Cc} and {\it I}4{\it cm} overlap at T $\simeq$ 100\,K and x $\simeq$ 0.495.

We have also undertaken  a  Rietveld analysis of neutron data on 
a Pb(Zr$_{0.49}$Ti$_{0.51}$)O$_{3}$ sample, collected at 10K at the ILL thermal source on the powder diffractometer D2B ($\lambda$=1.6 \AA). Analysis of the data collection from 2$\Theta$=20 to 140 degrees with a 2$\Theta$ step of 0.05 degrees  was carried out with the XND software \cite{XND}, using pseudo-Voigt peak-shape function including asymmetric broadening and linear interpolation for the background. The fluctuation in the sample's composition was estimated to be $\simeq$ 0.003 based on x-ray peaks at 300K and dielectric measurement.
We observed  superlattice reflections in the diffraction pattern (not shown here), and thus considered different space groups that are associated with doubled unit cells -- namely,  rhombohedral {\it R}3{\it c}, monoclinic {\it Pc}, {\it Cc} and {\it Ic}, and tetragonal {\it I}4{\it cm}. Each {\it single}-phase model leads to a considerable mismatch between the observed and calculated profiles for the pseudocubic reflections, and thus to poor agreement factors. A considerable improvement is achieved by considering {\it two}-phase coexistence models between the above space groups and also by including the monoclinic {\it Cm} and cubic ${\it Pm}\bar{3}{\it m}$ phases. The best agreement factors and matching between the observed and calculated profiles is obtained with the {\it I}4{\it cm}+{\it Cc} coexistence model (that results in  R$_{wp}$,  R$_{B}$ and G.o.F. factors being equal to 
6.69\%, 5.09\% and 1.72, respectively). Our structural refinement thus confirms the existence of the ``controversial'' {\it Cc} state (for which our structural parameters are very close to 
the ones of Ref. \cite{Ranjan2}) and of the presently-discovered {\it I}4{\it cm} structure (for which the $a$, $b$ and $c$ parameters are equal to 5.700(2), 5.700(2) and 8.329(6) \AA, respectively).
Note that the proportion of {\it I}4{\it cm} and {\it Cc} phases is approximately 7:3. The coexistence of  these two phases may be due to compositional fluctuation in the sample, which will make this coexistence consistent with the theoretical phase diagram of Fig.2. It may also take its origin from the presence of small grains in the grown ceramic, since Ref.   \cite{Zembilotov} predicts that 
the coexistence of different phases may be energetically favorable (over a single phase) in a  ceramic of fixed composition when the grain size is small. 
We also hope that our other predictions, including the ones about multiphase points, will be experimentally confirmed soon.

This work is supported by ONR grants N00014-04-1-0413, N00014-01-1-0600 and  N00014-01-1-0365, by NSF grants DMR-0404335, and by DOE grant DE-FG02-05ER46188.
We thank B. Noheda for useful discussions.

\newpage

\begin{figure}
\caption{Supercell average  $< {\bf u}>$ of the local mode vectors (Panels (a), (b) and (d)) and  AFD-related  $< {\bf \omega}>_{R}$ 
quantity (Panels (c) and (e)), as a function of temperature  in  Pb(Zr$_{0.52}$Ti$_{0.48}$)O$_{3}$. Panel (a) 
corresponds to predictions from the use of the alloy effective Hamiltonian method of Refs. \cite{PRLPZT} -- that neglects AFD effects. 
Panels (b) and (c) (respectively, Panels (d) and (e)) shows the results of the presently proposed method  with no applied 
pressure (respectively, when applying a pressure of $-$4.68GPa).}
\end{figure}

\begin{figure}
\caption{Phase diagram of Pb(Zr$_{1-x}$Ti$_{x}$)O$_{3}$ near its MPB, as predicted by the present scheme with an applied pressure of $-$4.68GPa. Symbols display the direct results of our simulations, while lines are guide for the eyes. Indices 1, 2 and 3 indicate the multiphase points. 
The uncertainty on the transition temperatures is typically around 13\,K, except close to the multiphase points 2 and 3 for which this uncertainty 
is around 3\,K.}
\end{figure}


\begin{references}
\bibitem{footnoteaccuracy} Ref. \cite{Noheda2002} indicates that the Ti composition, for which the  {\it P}4{\it mm}-to-{\it Cm} transition 
occurs at 300K is $\simeq$ 48\% while our scheme predicts that this happens for $x$ $\simeq$ 47.3\%. This slight disagreement 
may be due to the difficulty for the simulations to accurately reproduce the strong compositional dependency of the  {\it P}4{\it mm}-to-{\it Cm} transition 
temperature (see Fig.~2), or  that real samples have compositional inhomogeneities.
\end{references}
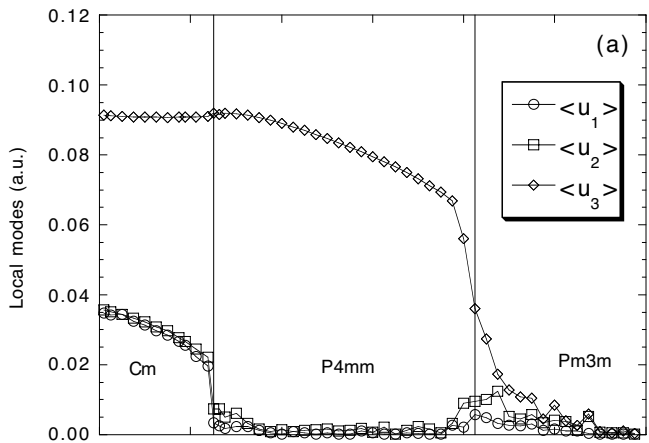
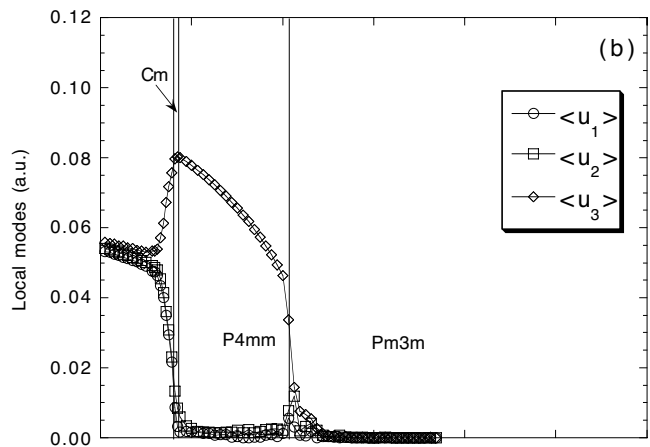
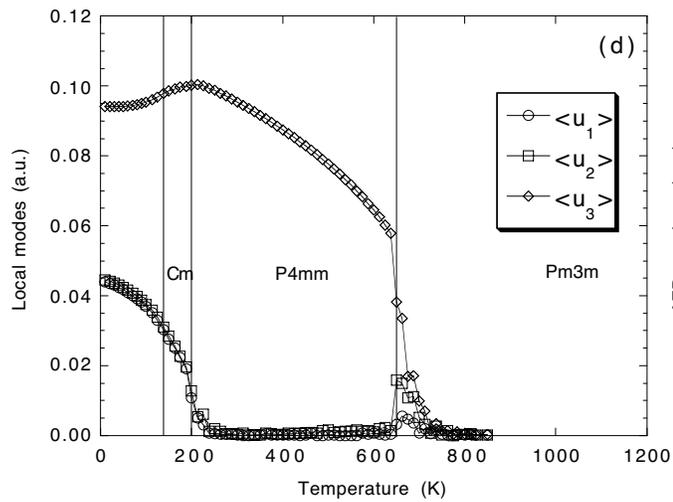
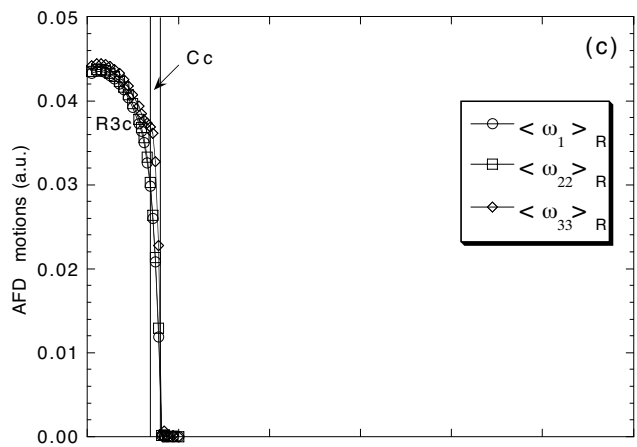
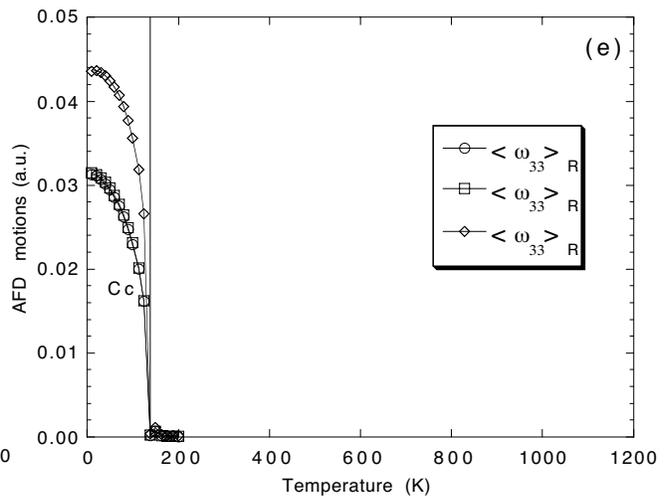
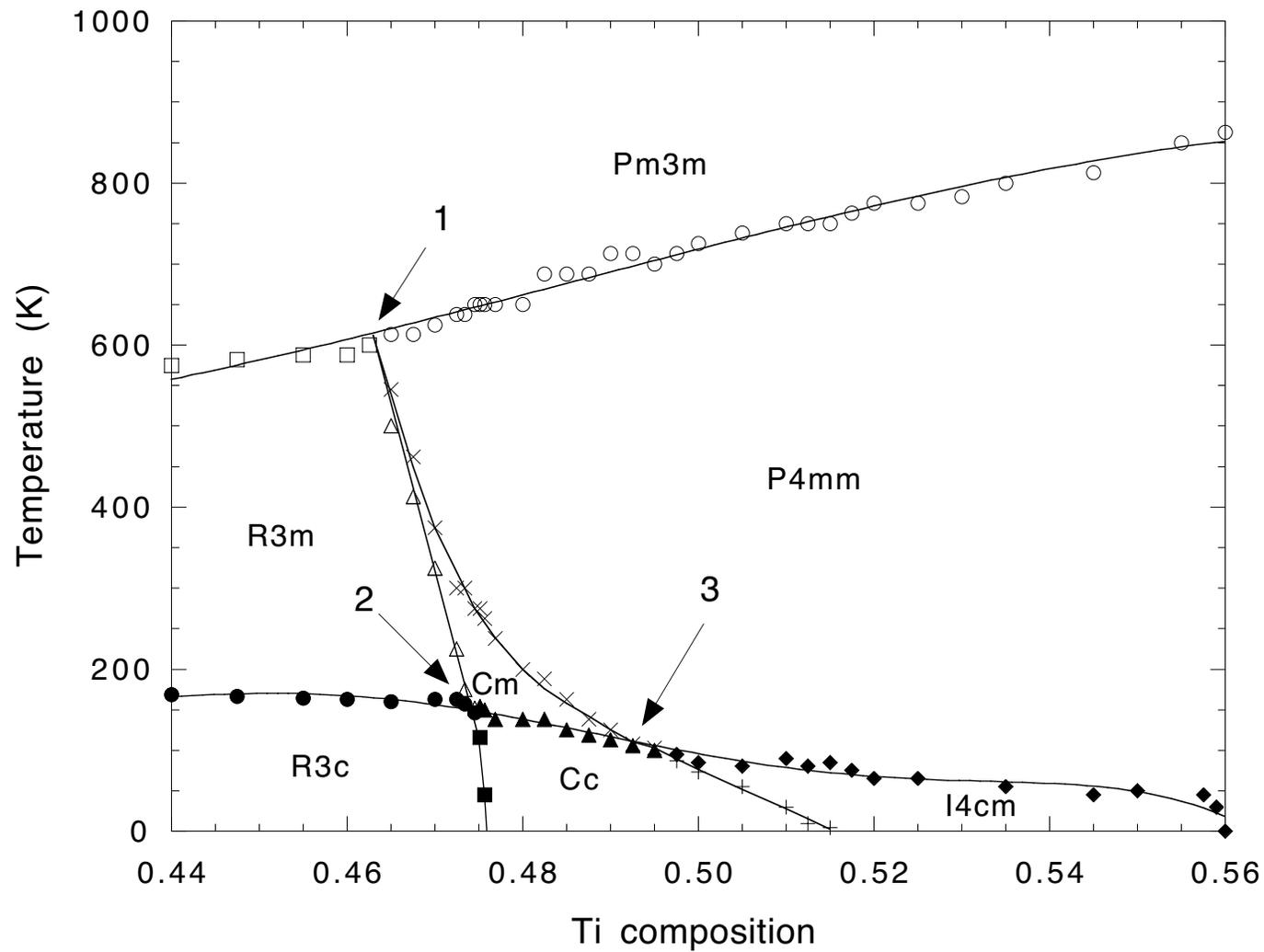
\end{document}